\documentstyle[12pt]{article}
\makeatletter
\@addtoreset{equation}{section}
\makeatother

\def\be{\begin{equation}}
\def\ee{\end{equation}}

\title{Nt-yet}
\begin{document}

\begin{flushright}
\vspace{1mm}
FIAN/TD/
 june {2002}\\
\end{flushright}

\vspace{1cm}

\begin{center}
{\large\bf TO PHYSICAL FOUNDATION OF QUANTUM MECHANICS }
\vglue 0.6  true cm
\vskip1cm
{\bf V.E.~Shemi-zadeh\footnotemark[1]} \footnotetext[1]{Internet adress: www.shemizadeh.narod.ru}
\vglue 0.3  true cm

I.E.Tamm Department of Theoretical Physics, Lebedev Physical Institute,\\
Leninsky prospect 53, 119991, Moscow, Russia
\vskip2cm
\end{center}

\begin{abstract}
\baselineskip .4 true cm
\noindent
On the base of years of experience of working on the problem of the physical 
foundation of quantum mechanics the author offers principles of solving it. Under 
certain pressure of mathematical formalism there has raised a hypothesis of 
complexity of space and time by Minkovsky, being significant mainly for quantum 
objects. In this eight-dimensional space and time with six space and two time 
dimensions all the problems and peculiarities of quantum mechanical formalism 
disappear, the reasons of their appearance become clear, and there comes a clear and 
physically transparent picture of the foundations of quantum mechanics.  
\end{abstract}

Over seventy-five years of quantum mechanics existence the physical essence of 
its foundations has remained unclear. Beyond the mathematical formalism of the 
theory there opens a strange physical picture contradicting common sense and logic. 
There is no doubt that the quantum mechanics is correct, as well as the results got by 
means of it. There has been offered a lot of interpretations, concepts and attempts to 
clarify the situation, but by now there is no satisfying clear concept [1]. 
Understanding of the foundations of the quantum theory remains today one of the 
actual problems. V.L. Ginzburg in his famous program article "What Problems of 
Physics and Astrophysics Seem Now to be Especially Important and Interesting" [2]. 
Mentions the problem of physical foundation of quantum mechanics among the three 
great problems . 
Conceptual failure of the theory leaves a feeling of it being incomplete in 
something very important. Such a situation is extremely unsatisfying for fundamental 
physical theory.
The present situation in the quantum mechanics bred a lot of paradoxes. We will 
consider one of them, simple and obvious, de Brogile paradox, following to [3].
In Paris there is a closed box, its inner walls are made of reflecting material. In 
the box there is a single particle. Without the attempt to localize the particle a 
partition wall is inserted into the box (this wall is also made of reflecting material), 
dividing it in two cells, which are the same. The cells are separated from each other, 
and one of the cells is sent to Tokyo . So the location of the particle is absolutely not 
determined. Where is the particle, in which of the cells?
The answer from the positions of the statistic mechanics is evident and obvious: 
the particle is either in the Paris cell or in the Tokyo cell. 
But from the positions of the quantum mechanics we have to admit that, in spite 
of the common sense, the particle is at the same time both in the Paris and the Tokyo 
cells! In spite of all the mystics of this, it is so! It is as true as the statement of electron 
s passing at the same time through both splits in the diffractional experiment!
Then in Tokyo there is carried out an experiment of finding the particle, where 
the particle is found, for instance, in Tokyo . Then at the same time the particle 
disappears from the Paris cell. So the experiment carried out in Tokyo causes an 
instant effect in Paris without any possible communication between the two cities. 
There is an authoritative point of view that this strangeness is in the nature of 
things and there is no need to think of its origin. As for me, personally I have not 
believed in these statements. I have always believed in principle possibility of making 
the unclear things and paradoxes of quantum theory clear. 
When I was a student I tried to understand physics, hiding itself behind the 
strangeness of mathematic formalism of quantum mechanics, making different simple 
and complex models, setting to different concepts. All attempts being failed I 
sometimes faced queer phenomena in the spirit of Wiegner's unconceivable 
effectiveness of mathematics in physical sciences when mathematical formalism 
seemed to suggest acceptance of the necessary concepts.
A few of these concrete cases were connected with the complexity of the 
quantum theory. 
Organic adherence of quantum mechanics and quantum theory of the field to the 
complex calculation is well known (complex analicity, dispersion relations, crossing- 
symmetry etc.) We should specially mention successfully developed Euclidean 
strategy in the Constructive field theory. 
The recent example is connected with our attempt to build quantum mechanics as 
non-local generalization of the classical mechanics . The sources of this work are 
simple and arise to analyses of Wiegner formulation of quantum mechanics in the 
phase space. 
For the Hamiltonian ${\cal H}$ and for the probability density $w(x,p,t)$
in the phase space Wiegner's equation of motion has the form

$$\frac{\partial w}{\partial t}=\frac{2}{\hbar}\sin \frac{\hbar}{2}
\left( \frac{\partial}{\partial p_w} \cdot \frac{\partial}{\partial x_H}-
\frac{\partial}{\partial p_H} \cdot \frac{\partial}{\partial x_w} \right)
{\cal H}w , \eqno(1)$$

where indices of both coordinates and impulses in brackets mean which of the 
functions ${\cal H}$ or $w$ is operated by the corresponding operator. 
Expanded, the Wiegner equation looks like this:

$$\frac{\partial w}{\partial t}+\{ {\cal H},w\} =
\frac{2}{\hbar}\sum_{k=1}^{\infty} (-1)^k\left( \frac{\hbar}{2}
\right) ^{n-1} \left( \frac{\partial ^n w}{\partial p^n}\cdot
\frac{\partial ^n {\cal H}}{\partial x ^n}-\frac{\partial ^n
{\cal H}}{\partial p ^n}\cdot \frac{\partial ^n w}{\partial x ^n}
\right) , \ \ n=2k+1$$

Without the right part it coincides with the Liuville equation of classical 
mechanics. Appearance of higher derivatives in the right part of the equation from the 
physical point of view can be interpreted as non-locality of the theory. Basing on this 
attempt we have considered non-local generalization of classical mechanics . Without 
going into details we shall point out the key moments. 
The very non-locality itself is inputted by means of stochastic quantum 
coordinates $\xi$ and impulses $\eta$:
$$x\to x+\xi, \ \ \ p\to p+\eta , $$

Where the average $<\xi >=0$, $<\eta >=0,$ and dispersion satisfies the relation
$$<\xi ^2><\eta ^2>=\hbar ^2/4. $$

Further on, introducing the non-local derivative according to the rule 
$$\frac{DF}{Dx}=\frac{F(x+\xi ,p)-F(x-\xi , p)}{2\xi }
 = \frac{1}{\xi } sh \left( \xi \frac{\partial}{\partial x}\right)
 F(x,p) \eqno(2)$$

considering the relations [7]: 
$$\xi w=\frac{\hbar}{2} \frac{\partial w}{\partial p}
\, \  \ \ \eta w=\frac{\hbar}{2} \frac{\partial w}{\partial x}, $$

we come to 
$$\frac{\partial w}{\partial t}=\frac{2}{\hbar} sh \frac{\hbar}{2}
\left( \frac{\partial}{\partial p_H} \cdot \frac{\partial}{\partial x_w}-
\frac{\partial}{\partial p_w} \cdot \frac{\partial }{\partial x_p}
 \right){\cal H}w , \eqno(3)$$

Looking very similar to (1) equation but having nothing to do with quantum 
mechanics.
This actually unexpected and somehow negative result played however a positive 
and decisive part further on.
Taking into account that $\sin x=-i\cdot  sh(ix)$, the difference between the equations 
of motion (1) and (2) can be eliminated under supposition of imaginarity of quantum 
coordinates and impulses, that is coming to complex values
$$t\to t+i\tau , \ \ x_k\to x_k +i\xi _k,\ \ k=1,2,3.$$.

And this in its turn requires complexification of space and impulse variables and 
transfer to 6-dimensional enclosing space. 
With imaginary $i\xi$ è $i\eta$ and non-local derivative (2) will take look of 
$$ \frac{DF}{Dx}=\frac{1}{\xi}\sin \left( \xi \frac{\partial}{\partial x}
\right) F(x,p).$$
and the motion equations (1) will transfer into Wiegner equation (3).
So it is turn out that we still can speak about building quantum mechanics on the 
base of classical mechanics, making radical supposition of the imaginarity of the 
quantum coordinates and impulses. And it means as it wasstaded before, the 
comlexification of Minkovsky spase, it's expansion from 4-dimensional to 8-
dimensional spase-time.
Summing up all the above-said and following the influence and hints of 
mathematical formalism we have accepted the following:

{\bf PROPOSITION:}
{\it Space and time are complexified manyfold (4 dimensional complex and 8-
dimensions material one). Additional (imaginary) dimensions are essential for 
micro-world and do not manifest themselves essentially for macroscopic objects.}

The above-said significantly clarifies all peculiarities of quantum mechanics, 
gives physical clarity and physical descriptions of the basic foundations of the theory. 
So, a micro-particle (electron), unlike microscopic objects , feels that it lives in 8-
dimensional space and time with six space and two time dimensions. 
A very important fact for us is that electron lives in two independent times $t$ è $\tau$.
Its coordinate $x$ for every fixed $t$ can take lots of values, corresponding to different time 
values $\tau $. That is from our point of view, from the positions of one-dimensional time, it 
will seem that electron at the same time is in different areas of space, in different 
mutually exclusive states. But now the reason of this seemed state is understandable: 
actually electron is in different states in different moments of the times $t_1=t+\tau_1, \ \ t_2=t+\tau_2$.
The coordinates values $x$, for instance of free electron for different 
values $\tau$ is distributed in space with some probability density 
$\rho (x,t)$ or amplitude $\Psi (x,t)$ which changes with time $t$.
 
Here again from the position of one-dimensional time it will seem that indivisible 
electron is in some strange way running aside forming some kind of a cloud. Now we 
know what takes place in reality. Now we can easily clarify the situation with paradox 
of de Brogile too, where the particle is at the same time in Paris and in Tokyo . Of 
course there is no similtaneousity, and it cannot be. The sameness of ordinary real 
times $t$ does not mean similtaneousity. The Paris and Tokyo particles (thus being the 
same particle) have different times $\tau$, so there is no paradox. 

Let us consider another question, the measure problem.
Mathematical formalism describes measurements as choice (projective postulat) 
from quantum states compositions$\Psi=\sum\psi_{i}$ and localizing (scattering 
et.cet.) one of it $\psi_{k}$ . All another states  at the same time are collapsing or changing in 
correspondig way $\psi_{k}\to \phi$. The measurement procedure implyes existens of macroscopic 
device. There is a natural question: what's the difference between macro device and 
microparticle?  Device consists of huge number $N$ of atoms and molecules. If square mean 
values of $\tau$ of consisting atoms $<\tau^{2}>$ 
then under assumtion of approximate equality all $\tau$ times for macrodevice 
time we can write for macroscopic device
\[
\tau_{Macro}\approx\frac{1}{N}\left(\sum_{i}<\tau_{i}^{2}>\right)^{\frac{1}{2}}
\approx\tau_{micro}/\sqrt{N}
\] 
Because for macroscopic device $N\approx 10^{23}$ , then $\tau_{macro}\approx 0$ in full accordance with 
hypothesis. 
That is the device "sees" the particle at the state with $\tau$ nearly zero 
and interaction with the particle measures it (localizing or scattering it) Another 
states at the instant are collapsing (decaying) In De Broigle paradox the 
state before measurement is superposition $\psi_{P}\oplus\psi_{T}$ .After measurement is done the 
state $\psi_{T}$ localizes $\psi_{T}\to\phi_{T}$ at the instant at the time $t$ , but in the time $t$ during the 
finite period of time $\tau$ is collapsing.

Thus under assumptions made above the second radical problem in foundation of 
quantum mechanic has the solution.

I have presented here only qualitative picture . It is not difficult to build 
corresponding mathematical formalism. 

The accepted Hypothesis looks rather verisimilar, especially after clarifying the 
physical foundations of the theory. It is not absolutely correct to believe I have 
introduced it into theory by force . As I have accentuated above, it appeared under 
pressure of mathematical formalism. If the Hypothesis turns out to be true, it must 
play a radical part for higher field theories and their latest generalizations.

The present work is performed with all-round support of the Center for 
Advanced Research:Theoretical and mathematical physics. The author thanks 
Smurniy E. D. for assistance to preparation of the article to publication. Also the 
author would like to thank Musin D.R. and Kutuzova T.S. for fruitful discussions.

\end{document}